\documentclass[aps,prl,twocolumn,superscriptaddress,showpacs]{revtex4-1}

\usepackage{graphicx}
\usepackage{color}
\usepackage{amsbsy,amssymb,amsmath,bm}
\usepackage{float}
\usepackage{epstopdf}

\begin{document}

\title{Quantum criticality and multiple crossing points in the magnetoresistance of thin TiN-films}

\author{K. Kronfeldner}
\affiliation{Institute of Experimental and Applied Physics, University of Regensburg, D-93025 Regensburg, Germany}
\author{T.\,I. Baturina}
\affiliation{Institute of Experimental and Applied Physics, University of Regensburg, D-93025 Regensburg, Germany}
\affiliation{Institute of Semiconductor Physics, 13 Lavrentjev Avenue, Novosibirsk, 630090 Russia}
\author{C. Strunk}
\affiliation{Institute of Experimental and Applied Physics, University of Regensburg, D-93025 Regensburg, Germany}
\email[]{christoph.strunk@physik.uni-regensburg.de}

\date{\today}

\begin{abstract}
We have measured $R(T,B)$ of a TiN thin-film very close to the disorder-driven superconductor-insulator transition but still superconducting at zero field and low temperatures.  
In a magnetic field we find three distinct crossing point of the magnetoresistance isotherms occur at magnetic fields $B_{cX}$ in three different temperature regions. Each crossing point in $R(T,B)$ corresponds to a plateau in $R(T,B_{cX})$. We systematically study the evolution of these crossing point near the disorder-induced superconductor/insulator transition, identify the most promising candidate for a quantum phase transition, and provide estimates for the two critical exponents $z$ and $\nu$.

\end{abstract}



\maketitle

Quantum phase transitions (QPTs) occur between competing ground states of many-body systems and result from quantum fluctuations near quantum critical points \cite{Sondhi_scaling}.
The superconductor/insulator transition (SIT) in thin metal films is viewed as one of the prime examples of a QPT \cite{Fisher_Grinstein, Fisher_scaling, Gantmakher_review}. It can be driven by one or more control parameters, such as the level of disorder in the system \cite{Haviland_onset} or the electron density $n$ \cite{caviglia}. More general, the relevant control parameter for the disorder driven SIT is $\Delta=k_F(n)\ell$, where $k_F(n)$ is the Fermi wave vector and $\ell$ the mean free path. For $\Delta\gtrsim\Delta_c$, the transition can also be driven by a perpendicular magnetic field $B$ \cite{Yazdani_1995}. In the limit of zero temperature, the discrimination between superconductor and insulator appears to be trivial, as the first case the resistance should be zero and in the other infinity. At the experimentally accessible finite temperatures, it has become common practice to regard the sign of the temperature derivative $\partial R(T,B)/\partial T$ of the resistance $R$ for the discrimination of the two ground states.

According to the theory of phase transitions, in a given universality class and near the critical point all properties depend in an universal way on a characteristic length $\xi\propto~|\Delta-\Delta_c|^\nu$ with the critical exponent $\nu$, and a characteristic frequency $\omega\propto\xi^{-z}$; $z$ being the dynamical critical exponent. If $R(T,B)$ is  controlled by the two critical exponents, the scaling relation $R(T,B)\propto|B-B_c|/T^{1/z\nu}$ is expected to hold in the vicinity of the crossing point \cite{Fisher_scaling}.  
 In a magnetic field distinct crossing points of the magnetoresistance isotherms $R(T=const.,B)$ at a critical magnetic field  $B_c$ were observed \cite{Hebard_1990,y_liu1993,Yazdani_1995, marcovic1998, gantmakher2000,Sambandamurthy_06}. Reversing Fishers' argument, the observed scaling property of $R(T,B)$ is often taken as a signature for critical behavior near a QPT and the crossing points were interpreted as evidence for a magnetic field induced SIT. 

The existence of a crossing point in $R(T=const.,B)$ implies that $R(T,B_c)=const$, i.e., a plateau in $R(T,B_c)$.  Experimentally, the latter condition is obeyed over finite $T$-intervals only. In addition,  more than one plateau was observed and interpreted als multistage criticality \cite{Popovic_Two_stage, Biscaras_2013}. On the other hand, plateaus in $R(T,B=const.)$ can be generated by two or more contributions of the resistivity with similar $T$-depencence, but opposite sign, leading to a cancellation of the $T$-dependence within a certain range. For example, the positive contribution of superconducting fluctuations to $R(T,B=const.)$ can be approximately compensated over a certain $T$-interval by the negative contribution from localization or DoS effects \cite{Baturina_Satta_2005, Baturina_Phase_Transitions, Tikhonov_2012}. 
Hence the basic question arises, how to ensure that an observed crossing point indeed  signals true quantum critical behavior.

\begin{figure}[b]
 \includegraphics[width=0.47\textwidth]{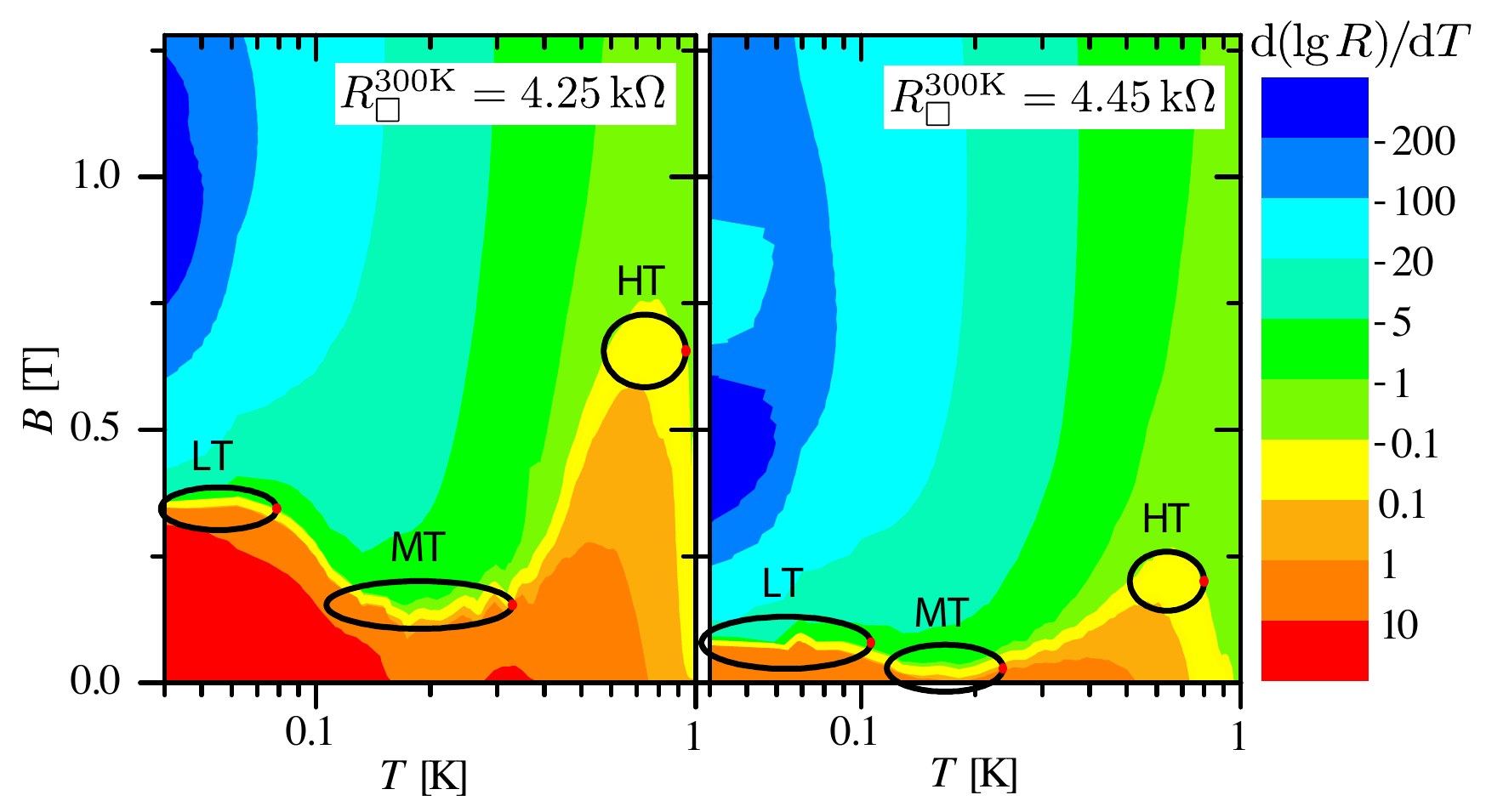}
\caption{\label{fig:Contour}
Logarithmic derivative of $R(T,B)$ vs.~temperature $T$ and magnetic field $B$. Blue/green areas indicate an insulating  ($\partial R/\partial T < 0$), red/orange areas a superconducting  ($\partial R/\partial T > 0$) trend. The nonmonotonic yellow strip displays the separatrix  $B^*(T)$ between superconducting and insulating regimes. The three extrema in $B^*(T)$ are marked with black ellipses labeled LT, MT and HT. The two panels display states with  different $R_{\square}^{300\mathrm{K}}$, i.e. different  levels of disorder (see text). 
}
\end{figure}

\begin{figure*}[t]
\includegraphics[width=.98\textwidth]{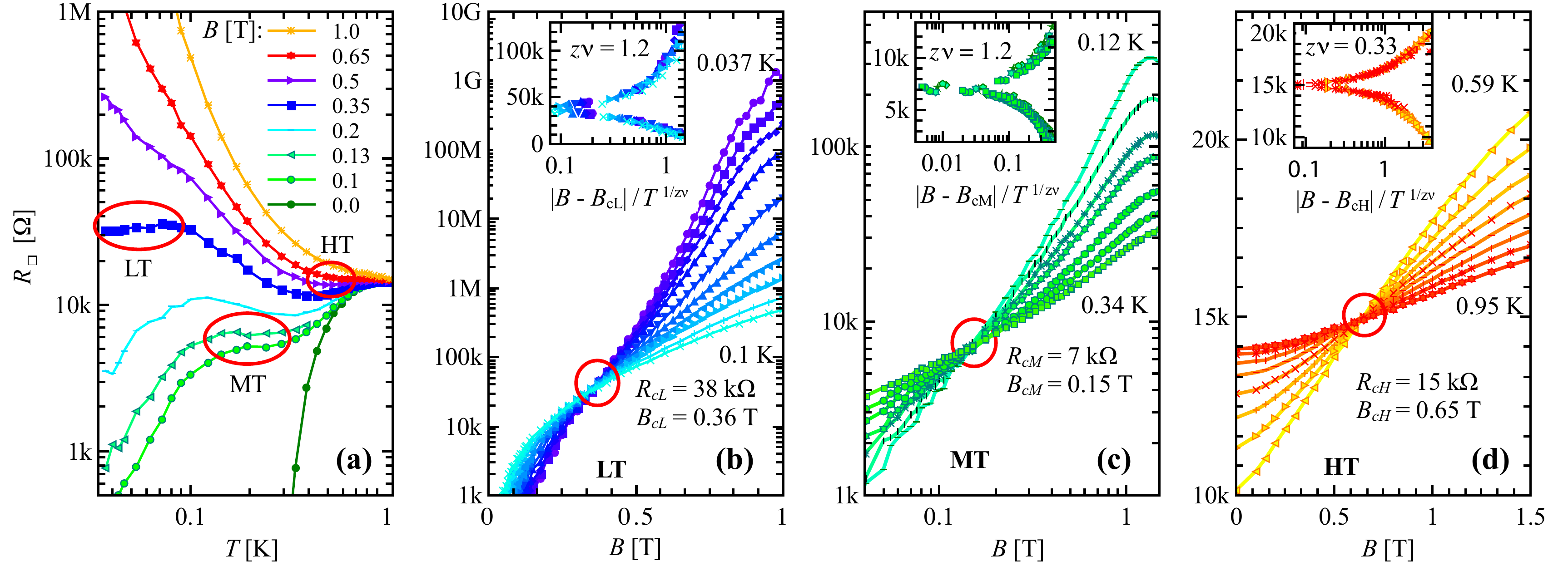}
\caption{\label{fig:RvB_four_pictures}
(a) Temperature dependence of thesheet  resistance $R_{\square}(T)$ for several magnetic fields and $R_{\square}^{300\mathrm{K}} = 4.25\,\mathrm{k}\Omega$. Three plateaus in the $R(T)$-curves at the extrema of $B^*(T)$ in Fig.~\ref{fig:Contour} are marked with red ellipses.  Lower inset: schematic of the TiN film with Au contacts. (b-d) crossing points in the $R(B)$-isotherms at the plateau regions in (a) corresponding to the extrema in  the separatix $B^*(T)$ (Fig.  \ref{fig:Contour}, left panel). 
 Insets: scaling analysis for each crossing point. Note that the horizontal scale in (b) and (d) is linear.
}
\end{figure*}

Here, we demonstrate that the scaling property is insufficient, as at least two control parameters are needed to establish a crossing point as a candidate quantum critical point for the magnetic field induced superconductor/insulator transition. In Fig.~\ref{fig:Contour} we show that the separatrix $B^\ast(T)$ (yellow strips), which separates regimes with positive and negative $\partial R(T,B)/\partial T$, can be strongly non-monotonic with three extrema of $B^\ast(T)$. Each extremum defines a $T$-interval displaying a distinct crossing of the $R(B)$-isoterms, i.e., a crossing point. We investigate the evolution of critical parameters with the level of disorder and argue that only one of the three crossing points remains as a candidate for the magnetic field-induced SIT. 

The approximately square shaped samples are patterned from the same $d \approx 3.6\,\mathrm{nm}$ thin TiN film (wafer D03 in \cite{Postolova2015}) with lateral size $L\geq 90\,\mu\mathrm{m}$ between two 100\,nm thick Au films as contacts on chip.  In this work we systematically tuned $R_{\square}^{300\mathrm{K}}$  towards the disorder-driven SIT by stepwise heating in air ($\approx 250^\circ$C for several minutes). The sheet resistance at room temperature  $R_{\square}^{300\mathrm{K}}\propto k_\mathrm{F}\ell$ provides a reliable measure for the level of disorder for different film thicknesses and different oxidation states \cite{supplement}. The current-voltage ($I$-$V$) characteristics are highly non-linear forcing us to always identify the linear regime of $I(V)$ to extract correct resistance values. We use a heavily filtered voltage biased 2-point circuit with an $I$-$V$ converter to measure $I(V)$ at different temperature and perpendicular magnetic field. The backaction noise from the $I$-$V$ converter was suppressed by a bias resistor.  The contact ($\simeq 6\,\Omega$) 
 resistance was subtracted.

The three extrema in $B^*(T)$ correspond to plateaus labeled LT, MT and HT in the $R(T)$ dependence  shown in Fig.~\ref{fig:RvB_four_pictures}a,  from which the data in Fig.~\ref{fig:Contour} is derived. The corresponding magnetoresistance isotherms at first sight do not reveal  well defined crossings. However, when we plot $R(B)$ in Fig.~\ref{fig:RvB_four_pictures}b-d separately  for the three plateau regions in Fig.~\ref{fig:RvB_four_pictures}a,   three distinct crossing points become clearly visible. In the Supplementary Material \cite{supplement} we propose empirical expressions for the functional form of the $R(T,B)$ curves.

Each of the crossing point allows for a scaling analysis \cite{Fisher_scaling,Gantmakher2003}. The result is shown in the insets of Fig.~\ref{fig:RvB_four_pictures}b-d. In the LT and MT regimes, we obtained a best data collapse of the $R(T,B)$ data with the scaling exponent $z\nu = 1.2\pm 0.2$, consistent with the findings in, e.g., indium oxide \cite{Hebard_1990,Shahar_breakdown} and \textit{a}-MoGe \cite{Yazdani_1995}. A smaller value of $z\nu = 0.33\pm0.03$ is found in the HT regime.  The fact that we observe several such points may even suggest multistage quantum criticality \cite{Popovic_Two_stage, Biscaras_2013}. This rises immediately the question regarding the nature of the different stages.\cite{Baturina_Satta_2005}.

\begin{figure*}[t]
 \includegraphics[width=1\textwidth]{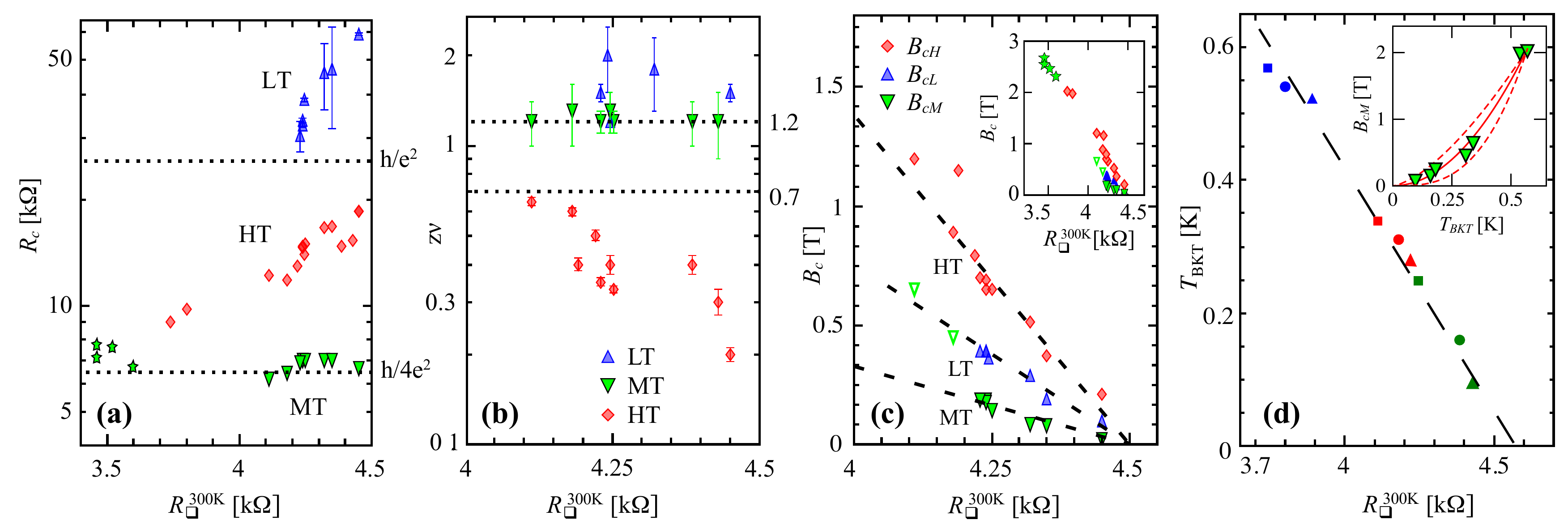}
\caption{\label{fig:Rc_Bc_zv_statistics}
(a) Critical resistance $R_c$ vs.~sheet resistance at room temperature $R_{\square}^{300\,\mathrm{K}}$. Green stars correspond to disorder levels, for which the MT- and HT-regimes are merged. ~(b) Scaling exponent $z\nu$ vs.~$R_{\square}^{300\,\mathrm{K}}$ resulting from the best data collapse of the $R(B)$-isotherms in the scaling plots (insets of Fig.~\ref{fig:RvB_four_pictures}b-d). The values $z\nu = 0.7$ and $z\nu = 1.2$ are marked by horizontal dotted lines. ~(c) Magnetic fields at the crossing points $B_{cL}$, $B_{cM}$ and $B_{cH}$ vs.~ $R_{\square}^{300\,\mathrm{K}}$. Dashed lines are guides to the eye. Inset: evolution of $B_{cL}$, $B_{cM}$ and $B_{cH}$ for a wider range of vs.~ $R_{\square}^{300\,\mathrm{K}}$. ~(d) BKT-transition temperature $T_\mathrm{BKT}$ extracted from $R(T,B=0)$ (see text) Inset: $B_{cM}$ vs.~$T_\mathrm{BKT}$. The solid red line corresponds to the exponent  $z = 1$. The dashed red lines are obtained for $z=4/3$ and $z=2/3$, respectively.
}
\end{figure*}

In order to identify crossing points as candidates for quantum criticality, we follow the evolution of the critical parameters $R_c$, $B_c$ and $z\nu$ as a function of the second non-thermal control parameter, i.e., the level of disorder, quantified by the normal-state resistance $R_{\square}^{300\,\mathrm{K}}$. We measured the $R(B)$-isotherms of TiN films with different oxidation states and perform the scaling analysis  for each of them. The evolution of the critical parameters vs. $R_{\square}^{300\,\mathrm{K}}$ is depicted in Fig.~\ref{fig:Rc_Bc_zv_statistics}.
Important features in Fig.~\ref{fig:Rc_Bc_zv_statistics} are the disorder independent values for $R_{cM} \approx h/4e^2$ (Fig.~\ref{fig:Rc_Bc_zv_statistics}a) and $z\nu \approx 1.2$ (Fig.~\ref{fig:Rc_Bc_zv_statistics}b) in the MT regime. The critical magnetic fields $B_{c}$ for all three crossing points nicely extrapolate to zero at  $R_{\square}^{300\,\mathrm{K}} \approx 4.5 \,\mathrm{k}\Omega$, which we associate with the critical level of disorder required for the disorder-induced SIT at $B=0$.
 According to Fisher  \cite{Fisher_scaling}, the transition field $B_c$ vanishes as $\xi^{-2}$ at the disorder-induced SIT at $B=0$. The  correlation length $\xi$ characterizes phase fluctuations in the SC phase, while the characteristic frequency $\omega$ sets the energy scale for the Berezhinski-Kosterlitz-Thouless (BKT) temperature for the vortex-BKT transition:  $T_{BKT}\propto\xi^{-z}$. This leads the relation $B_c \sim T_{BKT}^{2/z}$ \cite{Fisher_scaling}, which allows for an independent estimate to the dynamical critical exponent $z$. We extracted $T_\mathrm{BKT}$ from $R(T,B=0)$ for different films and different oxidation states \cite{supplement}. The result is shown in Fig.~\ref{fig:Rc_Bc_zv_statistics}d. It turns out that  $T_\mathrm{BKT}$ decreases nearly linearly with  $R_{\square}^{300\,\mathrm{K}}$ and extrapolates to zero at the very same critical value of disorder indicated by  $R_{\square}^{300\,\mathrm{K}}=4.5\,$k$\Omega$.
 In the inset
 we plot $B_{cM}$ vs. $T_{BKT}$  (inset in Fig.~\ref{fig:Rc_Bc_zv_statistics}a) and find $2/z\simeq2$, corresponding to $z = 1$. This value is associated with long-ranged Coulomb interactions \cite{Fisher_scaling}.

Unlike the MT-regime,   we observe disorder-dependent and thus non-universal values for $R_{c}$,  and $z\nu$ in the LT- and HT-regimes. In particular, both $R_{cL}$ and $R_{cH}$ increase with increasing level of disorder (see Fig.~\ref{fig:Rc_Bc_zv_statistics} left panel).
The critical exponent $z\nu$ in the HT-regime is significantly smaller than 1.3 and decreases from $z\nu
 \approx 0.7$ to $z\nu \approx 0.2$ over the investigated range of $R_{\square}^{300\,\mathrm{K}}$ (see Fig.~\ref{fig:Rc_Bc_zv_statistics} middle panel). This is consistent with earlier findings in TiN 
($z\nu=1$,~\cite{Baturina_Satta_2005}), but also for a-MoGe ($z\nu=1$,~\cite{Yazdani_1995}),
NbSi ($z\nu=0.67$,~\cite{aubin2006}) and \textit{a}-Bi ($z\nu=0.7$,~\cite{markovic1999}) typically in films with $R_c<h/4e^2$.
In the LT-regime, the values for $z\nu$ are scattered around a mean $z\nu = 1.6 \pm 0.3$.


We now compare the properties of our TiN-films in the MT-regime to earlier observations in amorphous indium oxide films \cite{Hebard_1990,Sambandamurthy_06,Shahar_breakdown}. Similar to our observations, values of the critical resistance $R_c\simeq h/4e^2$ and the scaling exponent $z\nu\simeq1.2$ are close to those observed in sufficiently resistive InO$_x$, 
 albeit also values of $z\nu\simeq 2.4$ were reported for InO$_x$ \cite{Steiner_Kapitulnik_2008, Breznay2016}.
%
%
 In both materials $R(T,B)$ shows an Arrhenius-like behavior in this regime, with a characteristic temperature that changes sign at $B_\mathrm{cM}$. This sign change indicates the transition from vortex motion to Cooper-pair transport, or superconducting and insulating behavior, respectively, as suggested in the self-duality picture of Fisher \cite{Fisher_scaling}.  In addition, the value of $\nu = 1.2$ satisfies the theoretically predicted lower bound of $\nu \geq 1$ for a SIT in a disordered 2D system \cite{Fisher_Grinstein,Fisher_scaling}.
 The $R(T,B)$ curves are consistent with a power law $(B-B_\mathrm{cM})^{\alpha(T)}$ for $R<R_c$ as for InO$_x$ \cite{Hebard_1985,Sambandamurthy_06,Shahar_breakdown}.  Deeper in the insulating regime, the behavior crosses over to an exponential dependence on magnetic field (see the upturn in the double logarithmic plot in Fig.~\ref{fig:RvB_four_pictures}c and the Supplementary Material) that breaks duality symmetry \cite{Shahar_breakdown}.

Our study provides first experimental evidence that the scaling exponent $z\nu\simeq 1.2$ for the magnetic field-induced SIT remains independent of the level of disorder up to the disorder-induced SIT at B=0.  
Since quantum phase transitions fall into certain universality classes, such a disorder-independent $z\nu$ substantiates the conclusion that the crossing point in the MT-regime can be associated with a quantum phase transition.
On the other hand, the disorder-dependence of $R_{cL}$, $R_{cH}$ and $z\nu$ in the LT- and HT-crossing points reveals a lack of universality and thus discourages their interpretation as QCPs. 

\begin{figure}[t]
 {
 \includegraphics[width=0.38\textwidth]{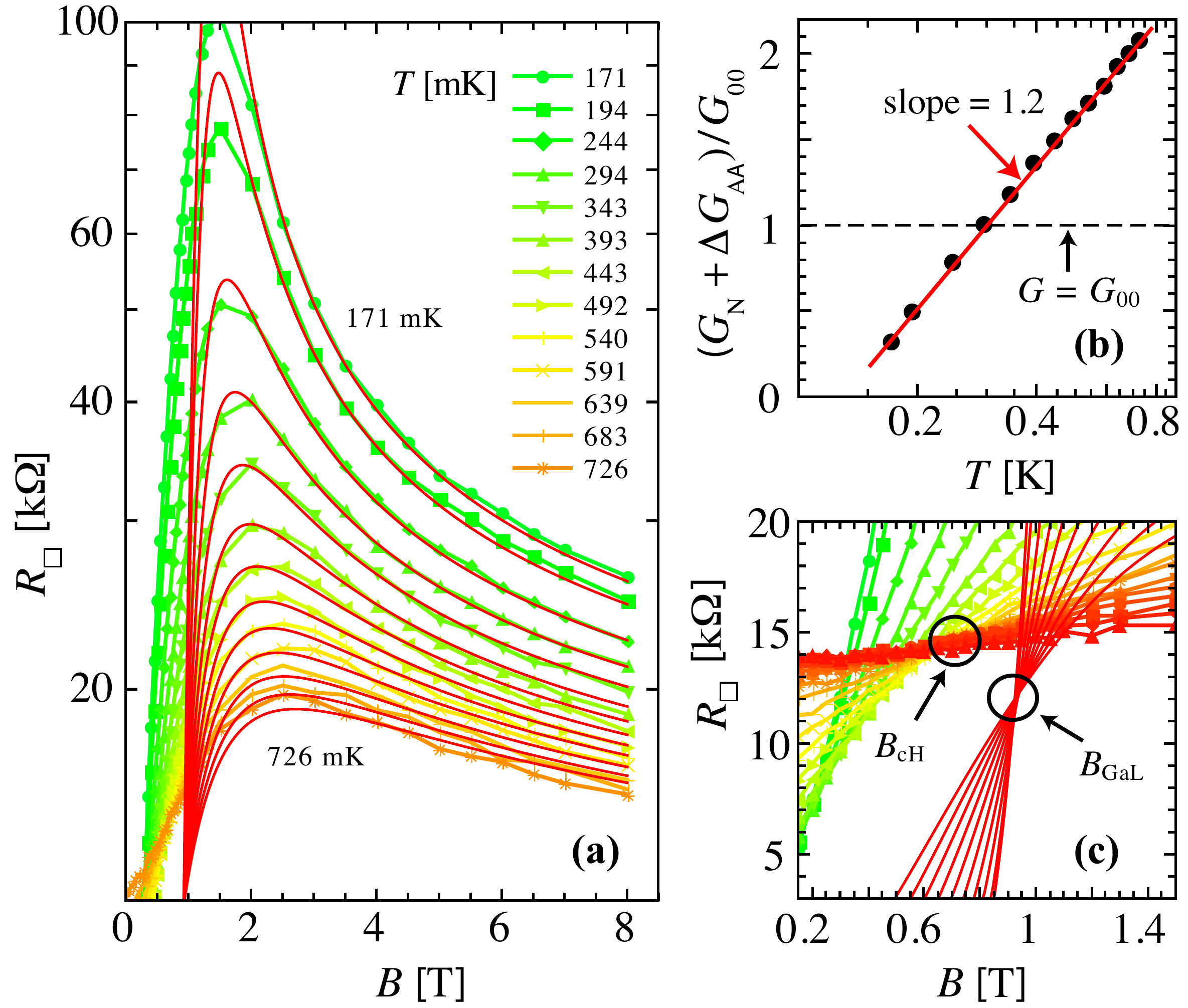}
 }
\caption{\label{Gal_Larkin_fit}
 (a) $R(B)$-isotherms for a sample with $R_{\square}^{300\mathrm{K}} = 4.228\,\mathrm{k}\Omega$. The red lines are fits of Eq.~S5 with the parameters $T_{c0} = 0.83\,\mathrm{K}$ and $B_{c2}(0) = 0.88\,\mathrm{T}$. (b) $G(T)$ extracted from the fits in the limit of large $B$. The red line denotes the Altshuler-Aronov contribution to $G$ (see text). (c) Zoom to the crossing points $B_{cH}  = 0.73\,\mathrm{T}$ and  $B_{GaL} = 0.935\,\mathrm{T}$.
}
\end{figure}

We now analyze the magnetoresistance in the HT-regime. For $B\gtrsim2\,$ the set of $R(B)$-curves in Fig.~\ref{Gal_Larkin_fit}a can be fitted with the Galitski-Larkin theory  of superconducting fluctuations in a magnetic field \cite{Galitski_2001}, with fixed values of the mean field transition temperature $T_{c0} = 0.83\,\mathrm{K}$ and upper critical field $B_{c2}(0) = 0.88\,\mathrm{T}$. For magnetic fields above that of the resistance maximum $B_\mathrm{max}$, the perturbative theory describes the data rather well, even for low conductances approaching $G_{00}= e^2/\pi h\simeq (81 \,\mathrm{k}\Omega)^{-1}$.  In order to adjust the theory to the high field limit, a $B$-independent offset is added that takes into account the normal state conductance $G_N$ plus the Altshuler-Aronov (AA) correction $\Delta G_{AA}(T)$ as the only fitting parameter \cite{supplement}. The $T$-dependence of the offset in this limit is shown in Fig.~\ref{Gal_Larkin_fit}b. Its $T$-dependence is logarithmic with a prefactor of $1.2\,G_{00}$, in agreement with the AA-correction.  At $B<B_\mathrm{max}$ the agreement is only qualitative. The theory underestimates the resistance, because it does not contain the flux-flow resistance below $B_\mathrm{c2}$. Nevertheless it reproduces the observed resistance crossing, although at a somewhat higher magnetic field $B=B_\mathrm{GaL}>B_\mathrm{c2}(T)$ (Fig.~\ref{Gal_Larkin_fit}c). In lower resistance samples $B_\mathrm{cH}$ and $B_\mathrm{GaL}$ converge \cite{Gantmakher2003,Baturina_Satta_2005,supplement}.

The above analysis strongly suggests that the HT plateau in Fig.~\ref{fig:RvB_four_pictures}a and the corresponding HT-crossing point results from an accidental cancellation of different contribution to the $T$-dependent resistance  in the interval $0.6\lesssim T\lesssim 0.95\,$K, i.e., the interplay between vortex dynamics and the combined Altshuler-Aronov and Galitski-Larkin corrections to $R(T)$ \cite{supplement}. 
As the crossing of $R(B)$-isotherms is inevitable in the presence of sufficiently strong superconducting fluctuations, we suspect that some of the earlier reports on such crossings may be understood in terms of the interplay of flux flow and superconducting fluctuations rather than evidencing quantum criticality -- in particular if the sheet resistance at the crossing is low compared to $h/4e^2$.

\begin{figure}[t]
 \includegraphics[width=0.48\textwidth]{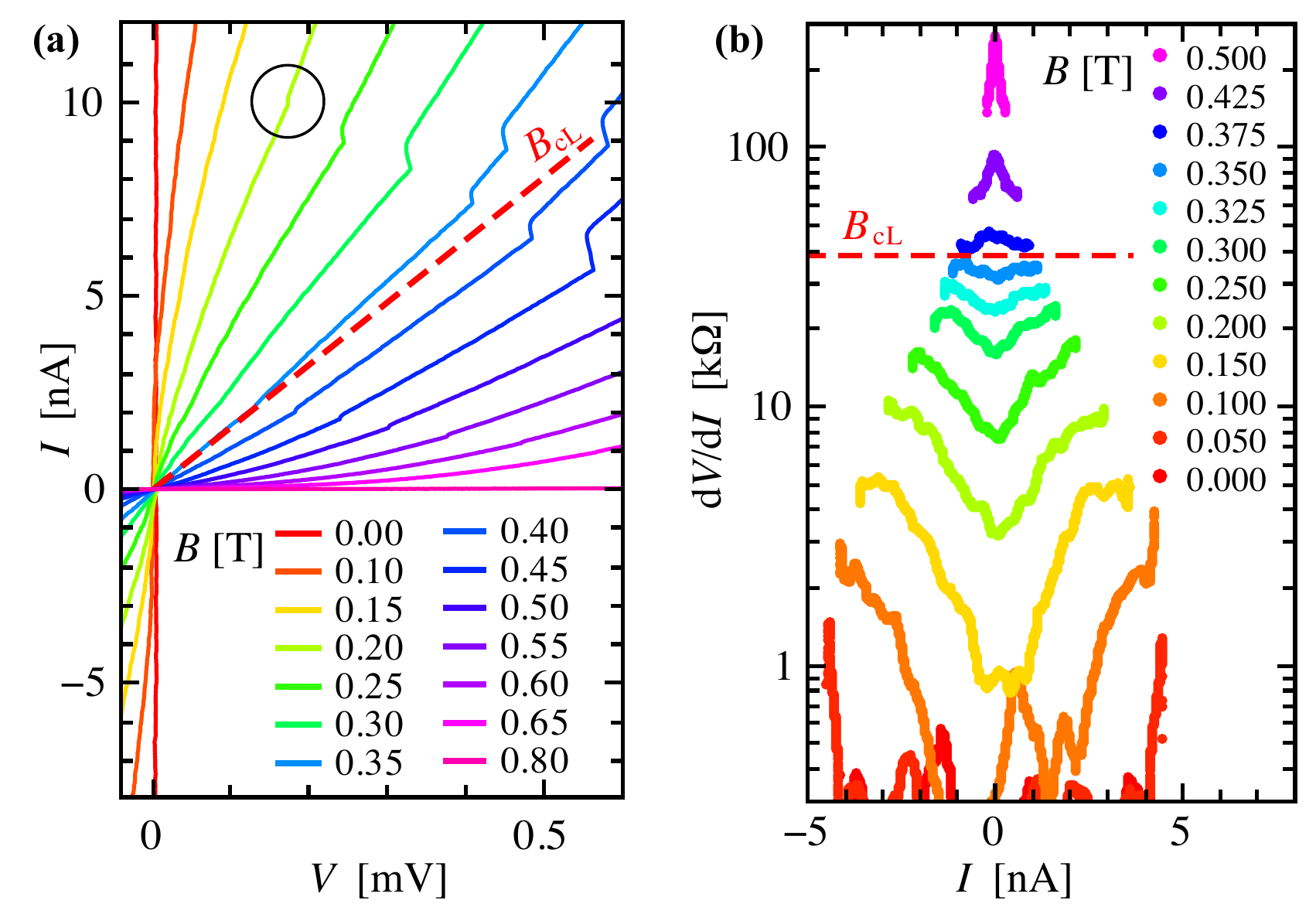}
\caption{\label{fig:IV}
(a) $I(V)$-characteristics for $T = 37 \,\mathrm{mK}$. The red dashed line indicates $B<B_{cL}$. The yellow line corresponds to the critical field $B_\mathrm{cM}$. The black circle indicates first appearance of a (tiny) jump in $I(V)$. The slight left tilt of the jumps results from an imperfect voltage bias, leading to a slight reduction of the voltage after the jump. (b) Corresponding differential resistance $dV/dI$ (right) for $I\lesssim 2\,\mathrm{nA}$. }
\end{figure}

The crossing point in the LT-regime also features non-universal $z\nu$ and $R_\mathrm{cL}>h/4e^2$. In order to elucidate its origin, we measured the highly non-linear $I(V)$-characteristics at low temperatures (Fig.~\ref{fig:IV}a). We observe current steps in the $I(V)$. These  occur for $B>B_\mathrm{cM}$ only, and are naturally explained as the breakdown of the insulating state along narrow low resistance channels.  Hence, they serve as a signature of the insulating state \cite{Baturina_localized,Cohen_2011} and confirm further that the SIT occurs at $B_\mathrm{cM}=0.15\,$T. On the other hand, in the low bias regime the differential resistance exhibits a crossover from a superconducting resistance dip to an insulating peak around zero bias right at $B_{cL}$ for increasing $B$ (Fig.~\ref{fig:IV}b). The crossing field $B_\mathrm{cL}$ is indicated by the dashed red lines in Figs.~\ref{fig:IV}a,b. The behavior at low bias and $B<B_\mathrm{cL}$ is consistent with that of a percolative network of superconducting filaments embedded into an insulating matrix close to the SIT \cite{Baturina_annals}, i.e., a dip in $dV/dI$ at low voltage.   

Between 0.25\,T\,$\leq B\leq$\,0.35\,T the resistance dip coexists with the current steps occuring at the same $T$ and $B$. This supports the coexistence of insulating and superconducting regions in the device.  For $B>B_{cL}$ their characteristics becomes insulating, i.e., $dV/dI$ develops a peak. 
At higher bias voltage, additional initially blocked current paths break through within the percolative network, leading to the observed steps in $I(V)$. Hence, our observations near the LT crossing point can be explained by an inhomogenous supercurrent distribution in which narrow conductive channels dominate the conductance at the lowest temperatures. Our finding implies that the behavior in the low-temperature limit can be misleading in the quest for quantum critical behavior.

In conclusion, we have shown that the mere possibility of scaling is insufficient to evidence quantum critical behavior in the vicinity of the SIT. The variation of a second control parameter is mandatory to demonstrate the universality of the  parameters controlling a presumed quantum critical point. We provide evidence the universality of a critical point in the magnetoresistance isotherms of thin TiN films with critical exponents $\nu=1.2$ and $z=1$.  Non-universal crossing behavior can be induced by inhomogeneity or the compensation of competing contributions to the conductance.

\begin{acknowledgments} We thank M.~Baklanov for the supply of TiN material, and V. Vinokur for inspiring discussions.
T. B. and V. Vinokur were supported by the A.~v.~Humboldt foundation.
\end{acknowledgments}

\bibliography{Kalok_TiN-bibliography}
\bibliographystyle{apsrev4-1}

\vspace{-4mm}

\clearpage

\pagebreak
\widetext
\begin{center}
\textbf{\large Supplemental Material for\\  "Crossing points and quantum criticality points in the magnetoresistance of thin TiN-films"}\\[3mm]
K. Kronfeldner,$^1$ T. I. Baturina,$^{1,2}$ and C. Strunk$^1$\\[3mm]
$^1$Institute of Experimental and Applied Physics, University of Regensburg, D-93052 Regensburg, Germany\\
$^2$ Institute of Semiconductor Physics, 13 Lavrentjev Avenue, Novosibirsk, 630090 Russia\\
\end{center}
\author{K. Kronfeldner}
\affiliation{Institute of Experimental and Applied Physics, University of Regensburg, D-93052 Regensburg, Germany}
\author{T.\,I. Baturina}
\affiliation{Institute of Experimental and Applied Physics, University of Regensburg, D-93052 Regensburg, Germany}
\affiliation{Institute of Semiconductor Physics, 13 Lavrentjev Avenue, Novosibirsk, 630090 Russia}
\author{C. Strunk}
\affiliation{Institute of Experimental and Applied Physics, University of Regensburg, D-93025 Regensburg, Germany}

\setcounter{equation}{0}
\setcounter{figure}{0}
\setcounter{table}{0}
\setcounter{page}{1}
\makeatletter
\renewcommand{\theequation}{S\arabic{equation}}
\renewcommand{\thefigure}{S\arabic{figure}}
\noindent

\section{Tuning the normal state properties near the d-SIT}
At room temperature the TiN-films are stable in air over years. However, when heated abover $\simeq 200\,^\circ$C in oxygen atmosphere or in air the sheet resistance gradually increases with a constant rate. This allowed us to study the same films at different levels of oxidation. It is not clear, whether the oxygen remains mainly at the surface, or is distributed more homogeneously within the film. The room temperature sheet resistance $R_{\square}^{300\mathrm{K}}$ constitutes a very reliable measure of the disorder strength.

\begin{figure*}[h]
 \includegraphics[width=0.825\textwidth]{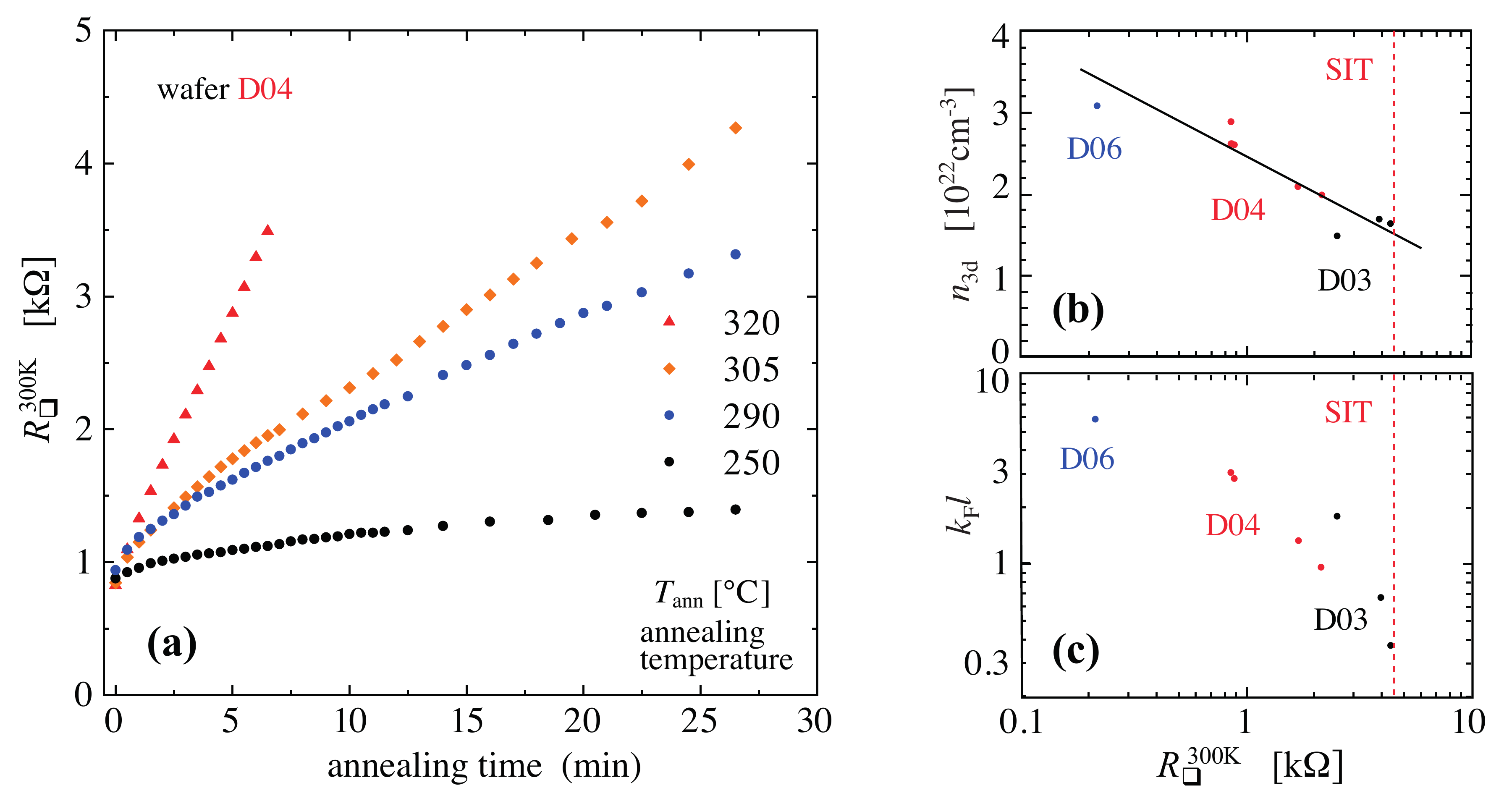}
 \caption{\label{density_TiN}
(a) Variation of the room temperature sheet resistance $R_{\square}^{300\mathrm{K}}$ over time in an oxygen atmosphere at different annealing temperatures $T_\mathrm{ann}$. ~(b) Electron density from measurements of the Hall effect at $T=10\,$K vs.~$R_{\square}^{300\mathrm{K}}$. ~(c) Corresponding $k_Fl$-product computed using the free electron model and an effective mass of $m^\ast\simeq 2m_{el}$. The vertical dashed lines indicate the position of the disorder-induced SIT.
}
\end{figure*}

In Figure~\ref{density_TiN}a we plot the electron densities of several TiN films with thicknesses of 10~nm (D06), 5~nm (D04), and 3.6~nm (D03) and several different oxidation times.  The electron density remains metallic, also in the vicinity of the d-SIT. The mean free path varies strongly between the films, and the $k_Fl$-product at the d-SIT is close to unity, as expected.

\begin{figure*}[h]
 \includegraphics[width=0.525\textwidth]{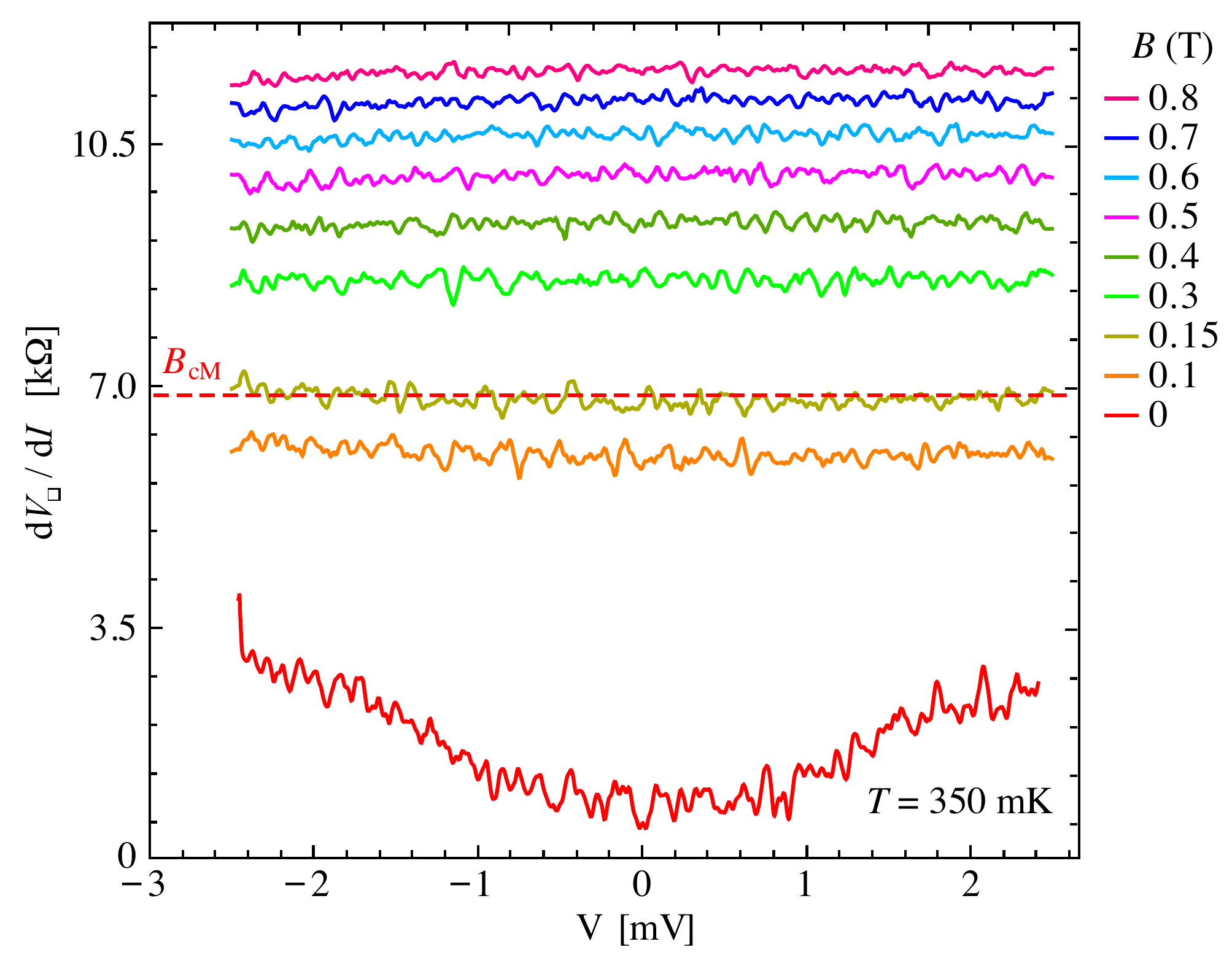}
 \caption{\label{MT_nonlinear}
Differential resistance vs.~voltage at $T=350\;$mK near the MT-crossing point. For voltages up to 1\;mV, the resistance is nearly linear, as opposed to the LT-regime (Fig.~\ref{fig:IV} in the main text).
}
\end{figure*}

\section{Non-linearity of $I$-$V$ characteristics}
The investigation of the low-temperature regime has revealed strongly nonlinear $I$-$V$-characteristics (see Fig.~\ref{fig:IV}). In Fig.~\ref{MT_nonlinear} we show that the resistance in the MT-regime is constant up to bias voltages $\simeq1\,$mV. The strongest nonlinearity occurs at zero magnetic field. This implies that crossing points in the MT-regime are much easier to capture, when compared to the LT-regime. Non-linearities in the HT-regime  are even less critical.

\section{Phenomenological expressions for $R(T,B)$}
In the following, we present phenomenological expressions that describe the measured $R(T,B)$ curves in a wide range of magnetic field, i.e., also away from the crossing points. In a wide range of magnetic field the behavior is no longer expected to be critical and universal -- nevertheless at the lowest $T$ a description of the $R(T,B)$-curves with very few parameters is still possible, while the range of collapse shrinks for higher $T$. 

\subsection{(i) LT-regime:} For $B_{cL}\leq B \lesssim 0.8\,\mathrm{T}$ the scaling behavior of experimental data can be parametrized by the  expression:
\begin{equation}\label{RvTB_LT}
R_{\square}(T,B) = R_{cL}\cdot \exp\left\{{A\cdot \left[{   \left({\frac{B}{B_{cL}}}\right)^{a_L(T)}  -1}\right] }\right\}
\end{equation}
where $B_{cL}=0.36\,\mathrm{T}$, $R_{cL}=38\,\mathrm{k}\Omega$, and the dimensionless constant $A\approx 2.4$ are kept fixed. The only adjustable parameter is the exponent $a_L(T)$ that controls the temperature dependence of $R_{\square}(T,B)$.  Note that the form of Eq.~\ref{RvTB_LT} is doubly exponential in $a_L(T)$. 

\begin{figure*}[t]
 {
 \includegraphics[width=0.975\textwidth]{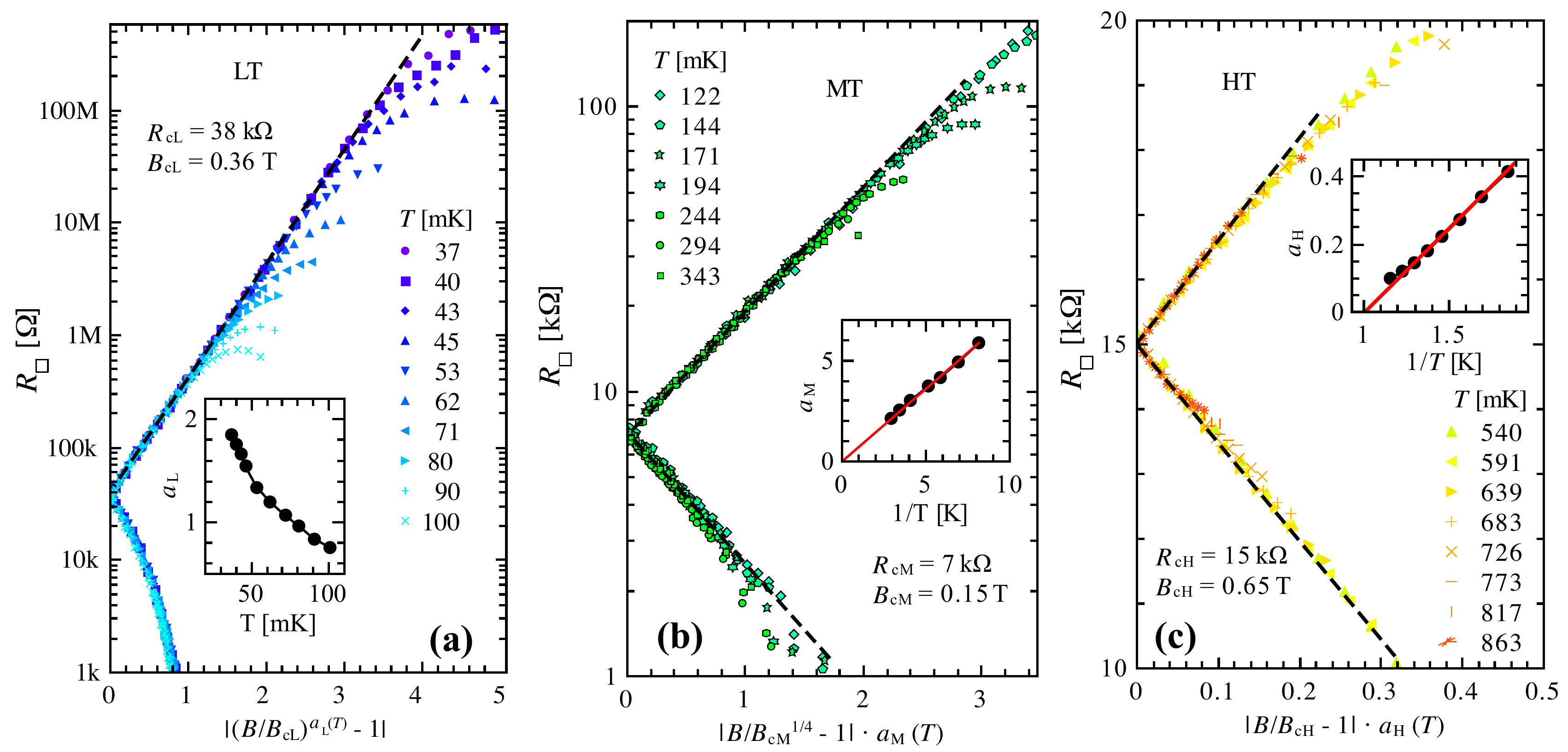}
 }
\caption{\label{RvB_scaling_alternative}
Scaling plots of the $R(T,B)$ (see Fig.~\ref{fig:RvB_four_pictures} in the main text), with the horizontal axes scaled according to the $B$- and $T$- dependences that are given by the phenomenological expressions \ref{RvTB_LT}, \ref{RvTB_MT} and \ref{RvTB_HT}. The dashed lines represent Eqs.~\ref{RvTB_LT}, \ref{RvTB_MT} and \ref{RvTB_HT}. The insets show the temperature dependence of the single fitting parameters $a_L(T)$, $a_M(T)$ and $a_H(T)$. The red lines in the insets are fits according to $a_M(T) = T_{0M}/T$ (panel b) which is associated with thermally acticated behaviour of $R(T)$ with activation energy $T_{0M} = 0.73\,\mathrm{K}$ in the MT-regime, and $a_H(T) = (T_{0H}-T)/2T$ (panel c) with $T_{0H} = 0.96\,\mathrm{K}$ in the HT-regime. 
}
\end{figure*}

 For $B<B_{cL}$, the scaled $x$-axis produces an excellent collapse of the $R(B)$-isotherms down to $R_\square\simeq 1\,k\Omega$.  For $B>B_{cL}$ the range of collapse increases with decreasing $T$: Eq.~\ref{RvTB_LT} overestimates the resistance with increasing $B$ as the magnetoresistance maximum is approached.  The exponent $a_L(T)$ increases from $\lesssim 1$ at $T = 100\,\mathrm{mK}$ up to almost $\approx 2$ at $T = 0.37 \,\mathrm{mK}$ (inset in Fig.~\ref{RvB_scaling_alternative}a). Sankar and Tripathi have predicted a very similar form of magnetoresistance with $a_L(T=0)=2$ when studying models of Josephson junction networks \cite{Sankar_Tripathy}. 


\subsection{(ii) MT-regime:} In the MT-regime, the scaling behavior of the $R(T,B)$-curves on both sides of the crossing point within $1\,k\Omega \lesssim R \lesssim 100\,\mathrm{k} \Omega$ can be described with the expression:
\begin{equation}\label{RvTB_MT}
 R_{\square}(T,B) = R_{cM}\cdot \exp\left\{{a_M(T)\cdot \left[{  \left({\frac{B}{B_{cM}}}\right)^{1/4}-1}\right] }\right\}
\end{equation}
where $B_{cM} = 0.148\,\mathrm{T}$ and $R_{cM} = 7\,\mathrm{k}\Omega$, with 
$a_M(T)$ as the only adjustable parameter. We find that the temperature dependence in the MT-regime is thermally activated, i.e.,~$a_M(T)=T_{0M}/T$ (inset of Fig.~\ref{RvB_scaling_alternative}b) with activation temperature of $T_{0M} = 0.73\,\mathrm{K}$, similar to  earlier observations  \cite{Sambandamurthy_Collective_Insulating,Baturina_localized}. 
  The peculiar exponent $1/4$ in the $B$-dependence allows to capture a deviation from simple power law behavior.  A power law was observed below $B_{cM}$ in InO$_x$, while above $B_{cM}$ deviations from it where attributed to an appearent breakdown of duality\cite{OvadiaKalok}. Below $B_{cM}$ our data are also consistent both a power-law behavior of $R(B)$ and with Eq. \ref{RvTB_MT}, if $R_{cM}$ is replaced by  $R_{cL}$ and $B_{cM}$ by $B_{cL}$.

\subsection{(iii) HT-regime:} Around $B_{cH}$ the magnetoresistance is much weaker, and $B(T,B)$ varies linearly with magnetic field (see Fig.~\ref{RvB_scaling_alternative}c):
\begin{equation}\label{RvTB_HT}
 R_{\square}(T,B) = R_{cH}\cdot \left[{1+ a_H(T)\cdot \left({  \frac{B}{B_{cH}}-1}\right) }\right]
\end{equation}
with $B_{cH} = 0.65\,\mathrm{T}$ and $R_{cH} = 15\,\mathrm{k}\Omega$. The temperature dependence of the $R(T,B)$ is again controlled by only one adjustable parameter, $a_H(T) = (T_{0H}-T)/2T$, where $T_{0H} = 0.96\,\mathrm{K}$ (see upper inset Fig.~\ref{RvB_scaling_alternative}c).

\section*{Superconducting fluctuations in the HT-regime}
\begin{figure}[tb]
 \includegraphics[width=0.75\textwidth]{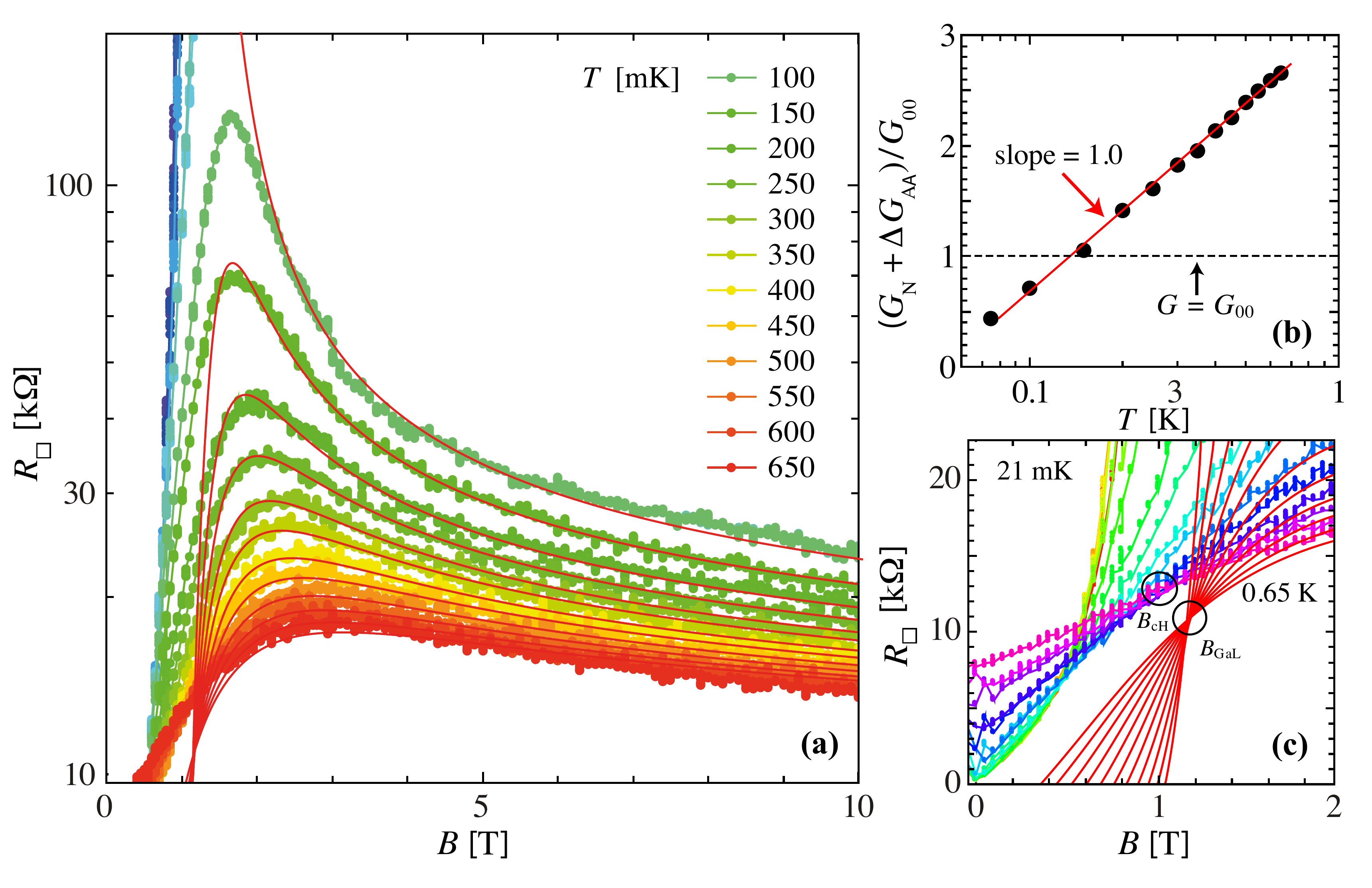}
\caption{\label{RvB_fit}
(a) $R(B)$-isotherms a different TiN-film  with $R_{\square}^{300\mathrm{K}} = 4.180\,\mathrm{k}\Omega$, i.e., slightly lower than the example in the main text. The red lines are fits of Eq. \ref{Galitski_total} to the $R(B)$-curves with the parameters $T_{c0} = 0.85\,\mathrm{K}$ and $B_{c2}(0) = 1.1\,\mathrm{T}$.  (b) $G(T)$ as extracted from the fits in the limit of large $B$. The red line denotes the Altshuler-Aronov contribution to $G$ (see text). (c) Zoom to the crossing points $B_{cH}  = 1.0\,\mathrm{T}$ and  $B_{GaL} = 1.08\,\mathrm{T}$.}
\end{figure}

Galitski and Larkin computed the fluctuation corrections to conductivity $\Delta G_{SF}$ above the upper critical field $B_{c2}$ in the low temperature limit \cite{Galitski_2001}.
Taking in to account the Aslamazov-Larkin, Maki-Thompson and density of states contributions for the dirty case where $k_BT_{c0}\tau/\hbar\ll1$  ($T_{c0}$ is the mean-field critical temperature and $\tau$ the scattering time), they obtained $\Delta G_{SF}$:

\begin{equation}\label{Galitski_delta}
 \Delta G_{SF} = \frac{2 e^2}{3\pi^2\hbar}\left [{-\ln{\frac{r}{h}}-\frac{3}{2r}+\Psi(r)+4 [{r\Psi'(r)-1}]}\right ]
\end{equation}

where $r = (1/2\gamma)h/t$, $\gamma = 1.781$ (Euler's constant), $h = (B-B_{c2}(T))/B_{c2}(0)$ and $t = T/T_{c0}$.  Tikhonov et al. \cite{Tikhonov_2012} extended Galitski and Larkin  result towards higher temperatures and magnetic  field.

Besides the fluctuation contribution, the conductivity contains the usual Drude term $G_N$ and a $T$-dependent offset that is independent of the magnetic field, which we attribute to the Altshuler-Aronov contribution from disorder-enhanced electron-electron interactions.  The total conductivity is then given by
\begin{equation}\label{Galitski_total}
 G(T,B) = G_N+\Delta G_{AA} + \Delta G_{SF}.
\end{equation}

The measured $R(B)$-isotherms (see Fig.~\ref{RvB_fit}a) can be fitted with equation \ref{Galitski_total}, starting above the $R(B)$-maximum up to the highest measured fields ($B_{max}=10\,\mathrm{T}$), for $R \lesssim (G_{00})^{-1} = (e^2/(2\pi^2\hbar))^{-1}\simeq 81\,\mathrm{k}\Omega$ (see also Fig.~\ref{Gal_Larkin_fit}c in the main text). The parameters $T_{c0}$ and $B_{c2}$ in Eq.~\ref{Galitski_total} were kept constant at $B_{c2} = 1.1,\mathrm{T}$ and $T_{c0} = 0.85\,\mathrm{K}$ in order to obtain the best fit of the $R(B)$-isotherms over the large  temperature range. We have used the Werthamer-Helfand result for $B_{c2}(T)$ \cite{Werthamer}.
The only free fitting parameter was $(G_N+G_{AA})/G_{00}$. It is plotted in Fig.~\ref{RvB_fit}b and behaves linear vs.~$\log T$ (see Fig.~4b in the main text). The prefactor of \mbox{$\approx 1$} is well consistent with the Aronov-Altshuler  correction \cite{Aronov_Altshuler}.
The red theory curves in Fig.~\ref{RvB_fit}a,c exhibit an approximate common intersection point at $B_{GaL}\approx 1.08\,\mathrm{T}$ and $R_{GaL} \simeq 11\,\mathrm{k}\Omega$. In the following we compare properties of this approximate crossing of fitting curves with the crossing point in the HT-regime.

\begin{figure}[t]
 \includegraphics[width=1\textwidth]{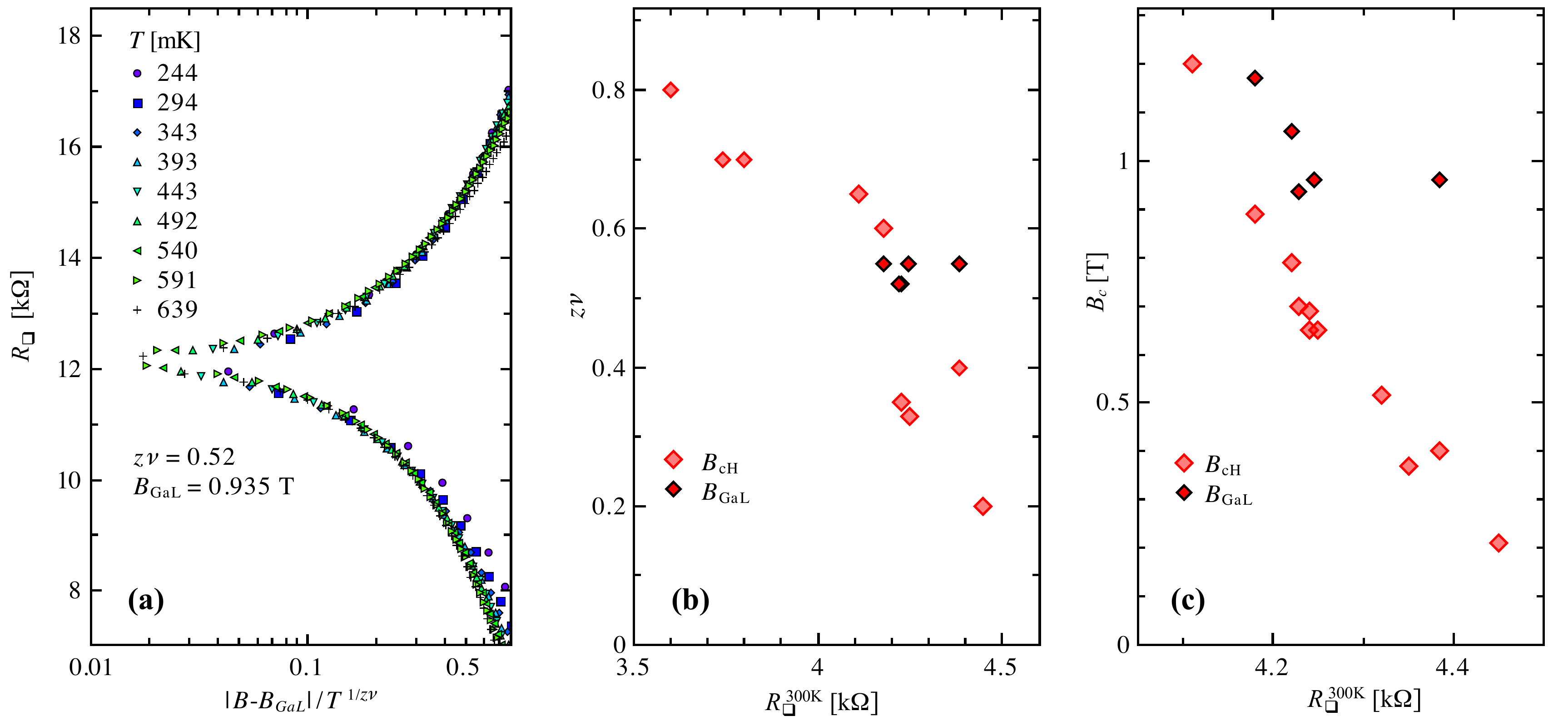}
\caption{\label{fig:scaling_Galitski}
(a) Scaling of the fitting curves from Fig.~\ref{Gal_Larkin_fit}a in the main text. The scaling exponent $z\nu = 0.52$ is determined by the best data collapse of the fitting-curves around $B_{GaL} = 0.935\,\mathrm{T}$. (b) Evolution of the scaling exponents $z\nu$, that are either determined from a scaling of the $R(T,B)$ data around the crossing point in the HT-regime (red diamonds) or from a scaling of the fitting-curves like in (a) (black diamonds with red core), vs. $B_{cH}$ and $B_{GaL}$, respectively. 
 (c) $B_{cH}$ (red diamonds) vs. $R_{\square}^{300\mathrm{K}}$ and $B_{GaL}$ (black diamonds with red core) vs. $R_{\square}^{300\mathrm{K}}$. Discrepancy between $B_{cH}$ and $B_{GaL}$ seems to vanish with decreasing disorder.
}
\end{figure}

\subsection{Scaling analysis of the Galitski-Larkin magnetoresistance} 
It was observed in earlier works \cite{Gantmakher2003, Baturina_Satta_2005} already that a  scaling analysis can be pursued at the approximate crossing point of the Galitski-Larkin magneto-resistance curves. In Fig.~\ref{fig:scaling_Galitski}a, we demonstrate scaling of the Galitski-Larkin magnetoresistance (red lines in Fig.~\ref{Gal_Larkin_fit}a in the main text) in a wide range $244\,$mK$\geq T \geq 640\,$mK.  The best collapse of data around the crossing field $B_{GaL} = 0.935\,\mathrm{T}$ was achieved with the scaling exponent $z\nu = 0.52$.

In Fig.~\ref{fig:scaling_Galitski}b we plot the resulting critical exponents $z\nu$ from the scaling of all crossing point in the HT-regime (red diamonds) together with those obtained from the scaling analysis of the fitting curves (black diamonds with red core) vs.~$R_{\square}^{300\mathrm{K}}$ (Fig.~\ref{fig:scaling_Galitski}). 
As described in the main text, the scaling exponent $z\nu$ decreases towards the D-SIT at $R_{\square}^{300\mathrm{K}} \approx 4.5\,\mathrm{k}\Omega$. The scaling exponents from the scaling analysis of the Galitski-Larkin theory fit rather well  into the experimental trace.

Furthermore, we can compare the crossing field   $B_{cH}$ of the magnetoresistance isotherms at the HT-crossing points
and the crossing points  $B_{GaL}$ of the corresponding fits. In Fig.~\ref{fig:scaling_Galitski}c we plot $B_{cH}$ (red diamonds) together with  $B_{GaL}$ (black diamonds with red core)  vs.~$R_{\square}^{300\mathrm{K}}$. While for $R_{\square}^{300\mathrm{K}}\simeq4.2\,$k$\Omega$ the values of $B_{GaL}$ rather well fit reasonably well again into the trace of $B_{cH}$, the point at $R_{\square}^{300\mathrm{K}}\simeq4.4\,$k$\Omega$ remains significantly above. As  $B_{GaL}$ is limited by the upper critical field $B_{c2}$, it seems that $B_{c2}$ remains nearly constant near the SIT at $R_{\square}^{300\mathrm{K}}\simeq4.5\,$k$\Omega$. 

We conclude that the HT-crossing point very near the SIT occurs at $B_{cH}<B_{c2}$ and below the Galitski-Larkin crossing at $B_{GaL}$. The Galitski-Larkin theory does not include the flux-flow resistance $R_\mathrm{FF}(T,B) = \Phi_0B\mu_V(T)$ that is expected for $B<B_{cH}$ in the absence of pinning. According to standard Bardeen-Stephen theory, the vortex mobility $\mu_V(T)$ is given by $\mu_V(T)=R_N/(\Phi_0B_{c2}(T))$, which {\it decreases} with decreasing $T$. Pinning is expected to further reduce $R_\mathrm{FF}(T,B)$ with respect to the Bardeen-Stephen value. In stark contrast, the experimental data in Fig.~\ref{fig:RvB_four_pictures}d show a linear, but much more moderate magnetoresistance, which extrapolates to $R(T,B=0)>0$ rather than zero. At magnetic field $B>B_{cH}$ the vortex mobility even  {\it increases} with decreasing $T$, corresponding to  negative $\partial R(T,B)/\partial T$. This behavior is inconsistent with the Bardeen-Stephen model and also pinning. The increase of the resistance above $R_N$ may be interpreted as a precursor of the long anticipated binding of charge/anti-charge pairs \cite{Fisher_scaling,Fazio_review,Mironov:2018aa}, very much dual to the binding of vortex-antivortex pairs at the Berezhinski-Kosterlitz-Thouless (BKT) transition. 

The flux-flow resistance below $B_{c2}$ must continuously cross over into the fluctuation resistance above $B_{c2}$.  It is thus not surprising that the position of the HT-crossing point is affected by the vortex contribution to the resistivity, and deviates from the Galitski-Larkin prediction with increasing disorder as the d-SIT is approached. Nevertheless, the similarities are striking. Together with the similarity of exponents and crossing fields (see Figs.~\ref{fig:scaling_Galitski}b,c) it is higgly suggestive to attribute the HT-crossing point to the combined effect of superconducting fluctuations and the localization, which mutually compensate each other in the HT-temperature regime.

\subsection{Berezhinski-Kosterlitz-Thouless transition near the SIT}
\begin{figure}[t]
 {
 \includegraphics[width=0.95\textwidth]{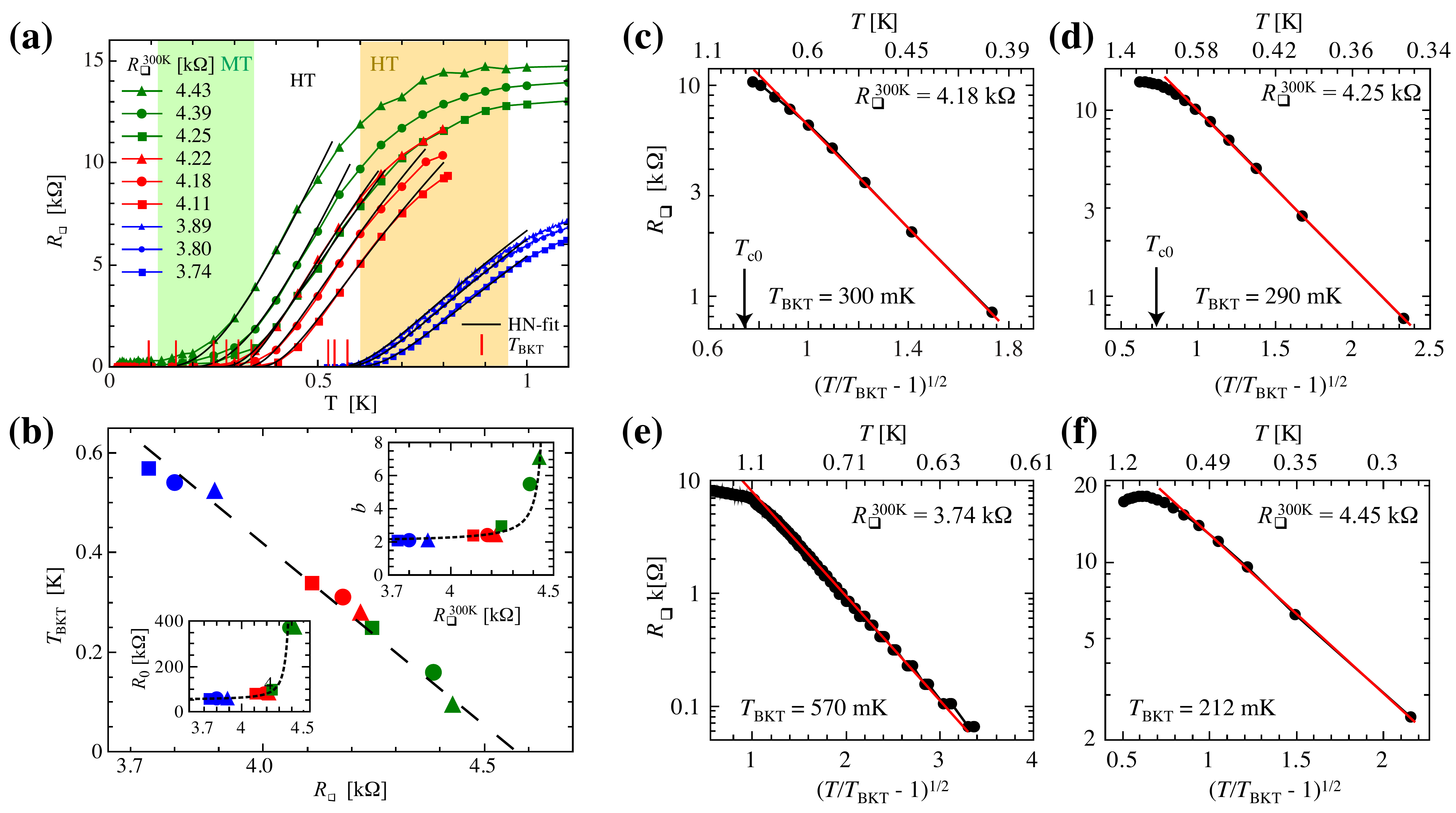}
 }
\caption{\label{RvT_HNfit_TBKT}
(a) Temperature dependence of the resistance at zero magnetic field for different levels of disorder $R_{\square}^{300\,\mathrm{K}}$. The colours denote subsequent oxidation states of the film. At each oxidation step, $R_{\square}^{300\,\mathrm{K}}$ increases by $200-400\,\Omega$. Films on the same wafer vary up to $\approx 200 \,\Omega$ in $R_{\square}^{300\,\mathrm{K}}$ at the same oxidation step. Black lines are fits to Eq. \ref{SRC}, with the fitting values for $T_{v-BKT}$ marked by the vertical red bars. ~(b) BKT-transition temperature $T_{BKT}$ at the D-SIT ($R_{\square}^{300\,\mathrm{K}}\approx 4.5\,\mathrm{k}\Omega$). Insets:  fit parameters $B$ and $R_0$ of (a).  Dashed lines are  guides to the eye. ~(c - f) Plots of $R(T,B=0)$ for several levels of disorder. The horizontal axis is chosen to display the Halperin-Nelson ('square-root cusp') behavior of Eq.~\ref{RvT_HNfit_TBKT} as straight lines.
}
\end{figure}

Mooij, Orlando and Beasley estimated in 1991 that with increasing disorder $T_\mathrm{BKT}$ is suppressed faster than the mean-field $T_{c0}$ \cite{Finkelstein_suppression,Mooij_Beasley}. In our significantly more disordered films this behavior becomes much more dramatic:
Figure~\ref{RvT_HNfit_TBKT}a shows $R(T,B=0)$ for several films with different degree of disorder. It is clearly seen that $R(T,B=0)>0$ in most of the HT-regime. When plotted vs. $\sqrt{T/T_\mathrm{BKT}-1}$ (Fig.~\ref{RvT_HNfit_TBKT}c-f) at linear dependence is revealed that is consistent with the Nelson-Halperin formula 
\begin{equation}\label{SRC}
 R(T) = R_0\cdot \exp\left[-\frac{b}{(T/T_{BKT}-1)^{-1/2}}\right]
\end{equation}
for the resistance  above the BKT-temperature $T_\mathrm{BKT}$,  where the thermal unbinding of vortex-antivortex pairs occurs.    The values of $T_{BKT}$ plotted vs.~$R_{\square}^{300\mathrm{K}}$ are show together with those of the parameters $R_0$ and $b$ in Fig.~\ref{RvT_HNfit_TBKT}b. It is seen that $T_{BKT}(R_{\square}^{300\mathrm{K}}$ decreases linearly and vanishes at $R_{\square}^{300\mathrm{K}})=4.5\,$k$\Omega$, just at the critical value for the d-SIT. The parameter $b$ is predicted to be of the order of unity, the values of $B\simeq 2$ have been observed before \cite{Postolova2015}.

When exposed to magnetic field, field-induced vortices are added to the thermally excited vortex-antivortex pairs and the vortex resistance increases towards the levels seen in Fig.~\ref{fig:RvB_four_pictures} (see main text).

The measured resistances $R(T,B=0)$ are shown in
Fig.~\ref{RvT_HNfit_TBKT}a.  The black lines are fits according to Eq.~\ref{SRC} with $T_{BKT}$  shown in Fig.~\ref{RvT_HNfit_TBKT}b together with parameters $b$ and $R_0$ depicted in the insets to Fig.~\ref{RvT_HNfit_TBKT}b. In accordance with the expectation, we find that $T_{BKT}$ indeed approaches zero at the D-SIT for $R_{\square}^{300\,\mathrm{K}} \approx 4.5\,\mathrm{k}\Omega$ \cite{Sacepe_disorder}. The Fisher relation  $B_{c}\propto T_{BKT}^{2/\nu}$ between $B_{cM}$ and $T_{BKT}$ allows for a determination of the dynamical critical exponent $z$ (see main text and the inset to Fig.~\ref{fig:Rc_Bc_zv_statistics}d).
%

\end{document}